\documentclass[preprint,12pt,sort&compress]{elsarticle}

\usepackage{amssymb}
\usepackage{inputenc}
\usepackage{lineno}

\biboptions{sort&compress}
\usepackage{microtype} % improves typography for PDF
\usepackage{booktabs, multirow} % for borders and merged ranges
\usepackage{soul}% for underlines
\usepackage[table]{xcolor} % for cell colors
\usepackage{changepage,threeparttable} % for wide tables
\usepackage[exponent-product=\cdot]{siunitx}
\usepackage[version=4]{mhchem}
\usepackage[breaklinks,colorlinks=true,bookmarks=false,citecolor=blue,urlcolor=blue]{hyperref} %latex w/dvipdf
\usepackage[capitalise]{cleveref}
\usepackage{svg}
\usepackage{makecell} %make headers bold

\DeclareSIUnit\angstrom{\text {Å}}

\begin{document}

\begin{frontmatter}


%% \affiliation{organization={},z

\title{Brightify: A tool for calculating directionally-resolved brightness in neutron sources}


\author[EPFL,JCNS]{Mina Akhyani} %\corref{cor1}
%\ead{mina.akhyani@epfl.ch}

\author[ESS]{Luca Zanini}
\author[EPFL]{Henrik Rønnow}

\affiliation[EPFL]{organization={Laboratory for Quantum Magnetism, Institute of Physics, École Polytechnique Fédérale de Lausanne (EPFL)},%Department and Organization
            addressline={Station 3}, 
            city={Lausanne},
            postcode={1015}, 
            country={Switzerland}}

\affiliation[JCNS]{organization={Jülich Centre for Neutron Science (JCNS)},
            addressline={Lichtenbergstr. 1},
            city={Garching},
             postcode={ 85748},
            country={Germany}}

\affiliation[ESS]{organization={European Spallation Source ERIC},%Department and Organization
            addressline={Partikelgatan 2}, 
            city={Lund},
            postcode={22484}, 
            country={Sweden}}

\begin{abstract}
Brightness is a critical metric for optimizing the design of neutron sources and beamlines, yet most Monte Carlo packages do not offer a direct, directionally-resolved calculation. Conventional tallies require predefined locations and directions, limiting their ability to capture localized features or non-uniform emission. In this paper, we present Brightify, an open-source Python-based tool designed to calculate brightness from Monte~Carlo Particle List (MCPL) files, which can be extracted from several Monte~Carlo simulation packages. Brightify provides an efficient computational approach to calculate brightness for any particle type and energy spectrum recorded in the MCPL file. It enables localized, directionally-resolved brightness evaluations by scanning across both spatial and angular domains, using the mean direction within each position window as a representative direction. This functionality is particularly valuable for identifying brightness hotspots and helping fine-tune the design of neutron sources for optimal performance. Validation against standard surface current and point estimator tallies confirms Brightify’s accuracy and demonstrates its added value in capturing features these methods miss. By addressing a methodological gap in brightness evaluation, Brightify offers a practical foundation for neutron source re-optimization, reduces computational burden, and accelerates source development workflows. The full code is available on the Brightify GitHub repository \cite{akhyani_mina_brightify_nodate}.

\end{abstract}

%\includegraphics{images/wp7_heat_dpdce.pdf}

%Research highlights -- submit as separate file
% \begin{highlights}
% \item Not part of editorial consideration and aren't required until the final files stage
% \item Only required for full research articles
% \item Must be provided as a Word document— select "Highlights" from the drop-down list when uploading files
% Each Highlight can be no more than 85 characters, including spaces
% \item No jargon, acronyms, or abbreviations: aim for a general audience and use keywords
% \item Consider the reader - Highlights are the first thing they'll see
% \end{highlights}


\end{frontmatter}


\section{Introduction}

In most neutron scattering experiments, a central design goal is to maximize the neutron flux on the sample. Achieving this involves both the efficient generation of neutrons at the target and their subsequent transport through neutron guides to the sample.
\\
Monte Carlo simulation codes such as PHITS \cite{niita_phitsparticle_2006}, MCNP \cite{forster_mcnp_1985}, OpenMC \cite{romano_openmc_2013}, Geant4 \cite{agostinelli_geant4simulation_2003}, and Fluka \cite{battistoni_overview_2015} provide powerful capabilities to model neutron production and transport. These tools enable the evaluation of various configurations taking into account different materials, geometries, and physical models. However, such simulations can be computationally demanding, especially when including detailed engineering models of complex source and instrument systems, which often makes full system simulations prohibitively time-consuming, even on high performance computing platforms \cite{lux_monte_2018}.

Simulating the complete neutron path, from the proton beam to the neutron reaching the sample, in a single step remains impractical due to the extreme inefficiency of the process: only about one neutron out of every half a million generated actually reaches the sample \cite{gallmeier_source_2005}. This inefficiency becomes especially problematic during iterative design processes, where numerous simulations are required. As a result, it is often more effective to decouple and optimize individual components of the system, such as the neutron source, transport to the guides, and transport to the sample. 

According to Liouville’s Theorem, the phase~space density of a system remains constant along trajectories in a conservative force field \cite{landau_lev_davidovich_course_1958}. In neutron transport, this principle implies that the brightness - which is directly related to the phase~space density — cannot increase as neutrons propagate through conservative systems. Between the neutron moderator and the sample, neutrons typically travel through free space without interactions or through neutron guides, where interactions are elastic and conserve energy. Consequently, the brightness at the sample cannot surpass that at the moderator surface and is normally lower due to the beam losses. This constraint means that the neutron flux deliverable to a confined phase~space volume at the sample is fundamentally limited, as dictated by Liouville’s Theorem.

In the last 30 years, substantial advances have been made in both the fabrication and design of neutron guides. These developments have been driven in part by the use of Monte Carlo ray-tracing tools such as McStas \cite{willendrup_mcstas_2014, lefmann_mcstas_1999}, which have facilitated more precise and efficient guide system modeling. Together, these improvements have made it possible to transport neutron phase space density with minimal loss, approaching the theoretical efficiency limit defined by Liouville’s theorem. As a result, attention has been turned towards optimizing the phase space density at the moderator exit. This shift effectively decouples the source optimization challenge into two separate domains: enhancing the source brightness, and improving the guide system that transmits neutrons from the source to the sample.

Current applicable methods for calculating brightness in Monte~Carlo simulations — such as surface current tallies and point estimator tallies \cite{niita_phitsparticle_2006, kulesza_mcnp_2022} — depend on the user to explicitely define both the location, and the direction in case of calculating angular neutron flux. This direction is typically fixed showing the direction toward the point estimator tally location or the normal direction to the surface, in case of surface current tally. This means that these techniques require a prior prediction of the scoring location and direction — essentially a guess about where in phase~space the brightness is likely to peak.

Whether or not this guess is valid depends critically on two factors: the underlying neutron distribution and the definition of the phase space volume over which brightness is evaluated. For sources with uniform or symmetrical neutron distributions, this guess is often trivial, and these methods perform adequately. However, when the source is non-uniform, predicting the region and direction of maximum brightness becomes much more challenging. Furthermore, in such non-uniform distributions, the size of the phase~space volume significantly influences the calculated brightness.

In particular, point tallies can only calculate brightness accurately when placed far from the source, such that the entire moderator surface lies within the corresponding phase~space volume and by using artificial collimators to define the required angular divergence \cite{zanini_design_2019, ornl_2017}. This is typically applicable in spallation sources and high-power research reactors, where neutron guides must be positioned several meters away from the moderator due to intense radiation levels.

In contrast, in compact low-energy neutron sources, the guides can be placed much closer to the moderator due to lower radiation levels \cite{bruckel_conceptual_nodate}. As a result, the required phase~space volume may cover only a portion of the moderator surface. In these situations, surface current tallies can be more practical for brightness calculations \cite{batkov_unperturbed_2013, rizzi_intense_2024}. However, because they calculate the neutron flux only in a single fixed direction \cite{niita_phitsparticle_2006, kulesza_mcnp_2022}, they cannot capture brightness in other angular directions. This is crucial in cases where the neutron emission is non-uniform and the peak brightness occurs in a direction different from the one assumed by the tally.

Therefore, neither surface current tallies nor point tallies are sufficient when a comprehensive scan of brightness across position and direction is required. The need for a more flexible and reliable tool was the motivation to create a tool for brightness calculation and it is the subject of this paper.

Brightify is a computational tool designed to scan a user-defined surface and identify the position and evaluate brightness at each position using the mean direction as a representative metric. It is particularly useful in cases where the phase~space region of interest is smaller than the full extent of the moderator area. Developed in Python, Brightify operates independently of any specific Monte Carlo transport code and uses the Monte Carlo Particle List (MCPL) format \cite{kittelmann_monte_2017} as its input. MCPL is a standardized binary format that captures detailed particle data, including type, position, direction, energy, polarization, and statistical weight \footnote{In Monte Carlo simulations, "weight" refers to the likelihood of an event occurring per primary particle. To obtain absolute intensity, this weight can be scaled by the known intensity of primary particles from the actual source.}. MCPL files are compatible with a variety of Monte Carlo simulation codes, and are designed to enable data sharing across platforms.

Because it is based on MCPL, Brightify can calculate brightness maps for any particle type and energy range contained in the input data. In the context of this paper, we focus specifically on neutrons. The tool outputs a directionally-resolved brightness map, highlighting how brightness varies with both position and direction. In another words, where to locate and how to align a neutron guide along representative directions to achieve higher brightness.

The structure of the paper is as follows: Section~2 provides the theoretical background, including the definition of brightness, an overview of Liouville’s theorem, and the relationship between the two. Section~3 reviews existing methods for brightness calculation and their limitations. In Section~4, the methodology and implementation of the Brightify tool are described. Section~5 presents a series of verification and validation examples demonstrating the application of Brightify and analyzing its performance. Finally, Section~6 offers concluding remarks and potential future upgrades.

\section{Theory: Brightness and Liouville's Theorem}

The state of a neutrons can be described by an 7-dimensional distribution function \( n(\vec{r}, \hat{\Omega}, E, t) \), where \( \vec{r} \) is the position vector, \( \hat{\Omega} \) is the direction of motion, \( E \) is the energy, and \( t \) is time. 

The number of neutrons per unit volume \( dV \), per unit solid angle \( d\Omega \), and per unit energy \( dE \), is commonly referred to as the neutron angular density. The brightness is then defined as the product of the particle speed \( v \) and the angular density. This quantity is also known as the angular flux, which is more commonly used in neutron transport theory to describe the directional distribution of neutrons. In mathematical terms:

\begin{equation}
        B(\vec{r}, \hat{\Omega}, E, t) = v \frac{n(\vec{r}, \hat{\Omega}, E, t)}{dV\,d\Omega\,dE} 
        = \frac{n(\vec{r}, \hat{\Omega}, E, t)}{dE\,dt\,d\Omega\,dA}
        \label{eq:general-brightness}
\end{equation}

where the infinitesimal volume is expressed as \( dV = dA\,dz = dA\,v\,dt \), assuming motion primarily along the z-direction. Eq.~\ref{eq:general-brightness} defines brightness as an instantaneous quantity.

Additionally, the phase~space density \( \rho(\vec{r}, \vec{p}, t) \) is defined as the number of particles per unit phase~space volume \( d\vec{r}\,d\vec{p} \). For a non-relativistic particle\footnote{ The non-relativistic assumption is valid for cold, thermal, and fast neutrons (meV to MeV range), where velocities are well below relativistic limits. Caution is needed when extending this to other particles near or above their rest mass energy.}, one can relate it to the angular density as:

\begin{equation}
        \rho(\vec{r}, \vec{p}, t) = \frac{n(\vec{r}, \hat{\Omega}, E, t)}{d\vec{r}\,d\vec{p}} 
        = \frac{v}{p^2} \frac{n(\vec{r}, \hat{\Omega}, E, t)}{dV\,d\Omega\,dE}
        \label{eq:general-ps-density}
\end{equation}

The factor \( p^2 \) in Eq.~\ref{eq:general-ps-density} arises from the Jacobian of the transformation from Cartesian to spherical coordinates in momentum space, and the energy differential \( dE \) follows from the kinetic energy relation \( E = p^2 / 2m \).

According to Liouville’s theorem, in the absence of dissipative forces, the phase~space density remains constant along particle trajectories. Therefore, under ideal conservative transport conditions, the brightness is conserved along trajectories, provided the momentum remains unchanged.

In a steady-state distribution, brightness becomes time-independent and is typically expressed per unit energy, area and solid angle. One may also define an energy-integrated brightness over a four-dimensional phase~space consisting of the transverse spatial and angular coordinates. This four-dimensional space is a projection of the full phase~space onto the transverse plane. In this context, the Eq.~\ref{eq:general-brightness} can be rewritten as the following:
\begin{equation}
        B_{E_2-E_1}(\vec{r}, \hat{\Omega})
        = \frac{1}{ \Delta\Omega\,\Delta A}\,\int_{0}^{\infty}dt\,\int_{E_1}^{E_2}dE\,\int_{\Delta \Omega}\,\int_{\Delta A}\,n(\vec{r}, \hat{\Omega}, E, t)
        \label{eq:4D-brightness}
\end{equation}

 If the neutron source distribution is uniform in both space and direction, the brightness remains constant regardless of the sampling region. However, if \( n(\vec{r}, \hat{\Omega}, E, t) \) varies spatially or directionally, then the computed brightness depends sensitively on the specific portion of phase~space being sampled. In non-uniform fields, changing \( \Delta A \) or \( \Delta \Omega \) can significantly alter the measured average of \( n \).

In practice, modeling neutron sources via Monte~Carlo simulations yields a discrete approximation of the neutron distribution. The accuracy of this representation depends on the number of sampled particles. Increasing this number reduces statistical error but significantly raises computational cost. 

Regardless of source uniformity, brightness in Monte~Carlo simulations must be computed over a finite volume \( \Delta A \cdot \Delta \Omega \). The choice of this volume often depends on the instrument's design and the beam transport system’s angular acceptance.

It is possible to analytically estimate the required viewable area on the moderator based on the instrument requirements, type of neutron transport system, the distance between moderator and the guide, etc \cite{konik_new_2023}.
One of the critical factors influencing \( \Delta A \) is the distance between the neutron source and the entrance of the guide system \cite{akhyani_mina_high_2024}. In high-power spallation sources and research reactors, this distance is typically large to reduce radiation damage, necessitating a larger \( \Delta A \). In contrast, compact accelerator-driven neutron sources (CANS) allow guide systems to be placed much closer to the moderator, enabling smaller sampling areas and improved spatial resolution.

For non-uniform sources, both \( \Delta A \) and angular sampling direction \( \Delta \Omega \) are crucial. A large \( \Delta A \) may include regions with minimal emission, reducing the effective brightness. Similarly, if \( \Delta \Omega \) is not aligned with the peak emission direction, the resulting brightness may significantly underestimate the source’s peak brightness. Normally, \( \Delta \Omega \) is constrained by the angular acceptance of the neutron transport system and is typically much narrower than the full angular distribution of the source. Therefore, to capture the peak brightness, it is essential that the calculation direction aligns with the peak emission direction. 

Accurate brightness evaluation, particularly for non-uniform sources, requires thoughtful selection of the sampling volume and careful alignment with the dominant emission direction.

\section{Methods for brightness calculation}

In this work, we focus on the calculation of neutron brightness using Monte~Carlo particle transport simulations. Among the available Monte~Carlo codes, MCNP \cite{forster_mcnp_1985} and PHITS \cite{niita_phitsparticle_2006} are two of the most widely used tools in the field. Accordingly, our review on the methods centers on the two built-in brightness estimation methods currently supported by these packages, as follows:

\begin{figure}[!htb]
\centering
\includegraphics[width=0.6\columnwidth]{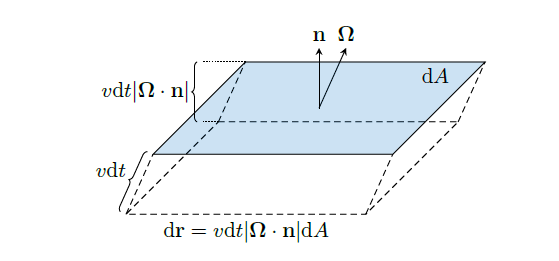}
\caption{Diagram for description of the surface current tally from MCNP 6.3.0 manual \cite{kulesza_mcnp_2022}.}
\label{fig:f1-cross}
\end{figure}

\begin{figure}[!htb]
\centering
\includegraphics[width=0.8\columnwidth]{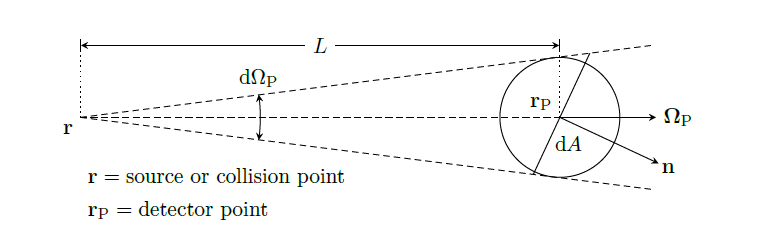}
\caption{Diagram for description of the point estimator tally from MCNP 6.3.0 manual. The point "r" can be from any place of the source \cite{kulesza_mcnp_2022}.}
\label{fig:f5-point}
\end{figure}

\begin{enumerate}
    \item \textbf{Surface current tally:} The surface current quantifies the number of particles, by their Monte Carlo weight, passing through surfaces that define cell boundaries, distributed across specified bins. The particle count at time $t$, within a volume element of $dr$, with directions within $d\Omega$, and energies within $dE$, is represented as $n(\vec{r},\hat{\Omega},E,t)\,dr\, d\Omega\, dE$.
    
    Consider a volume element $dr$ containing a surface element $dA$ (with surface normal $\hat{n}$) and extending along $\hat{\Omega}$ for a distance $\nu dt$, as illustrated in Fig.~\ref{fig:f1-cross}. The differential volume element is then determined as in Eq.~\ref{eq:dr}:
    \begin{equation}
    dr=\nu\, dt \left | \hat{\Omega} \cdot \hat{n} \right |\nu n(\vec{r},\hat{\Omega},E,t)\,d\Omega\, dE\, dt\, dA
    \label{eq:dr}
    \end{equation}
    
    Consequently, Eq.~\ref{eq:t-cross} indicates the count of particles crossing surface $A$ within energy bin $i$, time bin $j$, and angle bin $k$ \cite{kulesza_mcnp_2022}:
    \begin{equation}
    \int _{E_i}dE\int_{t_j}dt\int_{\Omega_k}d\Omega\int dA\left | \hat{\Omega} \cdot \hat{n} \right |\nu n(\vec{r},\hat{\Omega},E,t)
    \label{eq:t-cross}
    \end{equation}

    In MCNP and PHITS, this is typically computed using F1 and T-cross tallies. To derive brightness from such tallies, one must normalize the result by the surface area $dA$, and the solid angle $d\Omega$.

    A key limitation of this method is that the solid angle is always referenced with respect to the surface normal. As a result, when the direction of maximum brightness deviates from the normal, this approach can miss the true peak and underestimate brightness.

    This method is typically applied when the moderator is fully enclosed by a reflector, with no beam extraction channels—a setup known as “unperturbed brightness” \cite{batkov_unperturbed_2013}. While it can also be used in perturbed configurations, where openings exist for neutron extraction \cite{rizzi_intense_2024}, achieving accurate results in such cases typically demands significantly more computational effort. In these geometries, point estimator tallies are often preferred.

    \item \textbf{Point estimator tally:} 
    This method provides a deterministic estimate of particle flux at a specific location in space, based on contributions from current event point. Eq.~\ref{eq:t-point} shows what tally scores:
    \begin{equation}
         \frac{W \cdot p(\Omega_{p}) exp(-\lambda)}{L^2}
         \label{eq:t-point}
    \end{equation}
    where $W$ signifies the particle weight, $p(\Omega_{p})$ is the probability density for scattering in the direction $\Omega_{p}$ towards the point detector (assuming azimuthal symmetry). $\lambda$ represents the total mean free paths from the particle location to the detector, and $L$ is the distance from the source or collision event to the detector. Fig.~\ref{fig:f5-point} illustrates the contribution to the point estimator tally.
    
    This corresponds to F5 tally in MCNP and T-point in PHITS. This tally can be used only when openings are present in the source for neutron extraction, i.e. the perturbed brightness. 

    Based on its definition, point estimator tally records the contributions in 4$\pi$, therefore for brightness calculation, definition of artificial collimators are required to isolate the line of sight according to the required angular divergence \cite{zanini_design_2019, bruckel_conceptual_nodate}. Therefore, point detectors are placed sufficiently far from the moderator to have enough space for collimator definition. The measured flux—once normalized by $\Delta A / L^2$, with $\Delta A$ being the viewed area (typically the moderator surface)—is interpreted as the brightness in a specific direction and location. 

    It is important to note that this method only records neutrons and photons, and because it is an estimator, its accuracy depends on the nuclear data libraries. Consequently, it becomes unreliable above the energy limits of these libraries (typically above 20 MeV or 150 MeV, depending on the dataset).

    An advantage of the point estimator method is its computational efficiency, since it does not require tracking particles all the way to the tally location, in contrast to the surface current tally. However, it shares a key limitation with surface current tallies: if the point detector is not aligned with the location and direction of maximum brightness, it will fail to capture the peak.
\end{enumerate}

In summary, the surface current and point estimator tallies are valuable tools for brightness estimation when the location and direction of peak emission are predictable. This typically applies to uniform and geometrically symmetric sources.

However, in the case of non-uniform and non-symmetric sources, especially where instrument requirements demand high spatial resolution—meaning short source-to-guide distance where $\Delta A$ can be smaller than the moderator surface area—these conventional methods fall short. They lack the flexibility to scan the entire moderator surface across both spatial and angular dimensions to identify local brightness maxima.

This is where classical approaches reach their limit, and we need a more flexible tool to resolve local peaks and optimize instrument alignment—especially in complex source geometries where intuition alone is insufficient.

\section{Methodology of Brightify}

Brightify is an independent Python-based computational tool designed to flexibly search the space to find practical locations and representative directions for assessing brightness.

It operates on Monte~Carlo Particle List (MCPL) files~\cite{kittelmann_monte_2017}, a standardized binary format that records detailed particle state information (position, direction, energy, weight, etc.) and facilitates exchange between different Monte~Carlo simulation packages. MCPL files can be generated via conversion from simulation outputs such as SSW files (MCNP) \cite{kulesza_mcnp_2022} or dump files (PHITS) \cite{niita_phitsparticle_2006}.

In an ideal case, generating a complete brightness map would involve evaluating brightness for every individual neutron direction. However, this is computationally impractical, as MCPL files typically contain millions of neutron trajectories. As a result, a reference direction must be chosen for brightness calculations. In conventional tallies, this reference is typically the normal vector of the tally surface. Yet, relying on a single reference direction to represent such a large and diverse dataset often results in imprecise brightness estimates.

Brightify addresses this limitation by evaluating brightness across multiple representative directions rather than for each neutron individually or along just one axis. In essence, Brightify can be seen as a generalized version of the surface current tally that computes brightness for different directions.

The core idea is to divide the recording surface (flat or curved) into discrete position windows ($\Delta A$) and, within each, calculate brightness over a represntative direction. Brightify proceeds as follows:

\begin{itemize}

    \item Filter neutrons within each position window ($\Delta A$) on the recording surface.
    \item Determine the mean direction vector of the filtered neutrons, which is calculated by Eq.~\ref{eq:meanDir}:
        \begin{equation}
        \overline{V} = 
\frac{\sum_{i=1}^{n}\overrightarrow{V_{i}}\cdot w_{i}}{\sum_{i=1}^{n}w_{i}}
        \label{eq:meanDir}
    \end{equation}
    in which $w_{i}$ is the weight and $\overrightarrow{V_{i}}$ is the direction vector $(V_{x},V_{y},V_{z})$ of the $i$-th neutron in the position window.
    \item Filter neutrons within each position window that also fall within the specified angular divergence ($\Delta \Omega$) from the mean direction. 
    \item Compute brightness, in accordance with Eq.~\ref{eq:4D-brightness}, by summing the weights of selected neutrons and normalizing by the phase~space volume, as indicated by Eq.~\ref{eq:brightness}: 
    \begin{equation}
        B = \frac{\sum_{i=1}^{n} w_{i}}{N_{p}\times \Delta \Omega \times \Delta A }
        \label{eq:brightness}
    \end{equation}
    where $N_{p}$ is the number of total source particles, $\Delta \Omega$ is the solid angle, calculated by Eq.~\ref{eq:omega}:
    \begin{equation}
    \Delta \Omega = \int_{0}^{2\pi }\int_{0}^{\theta }sin\theta d\theta d\varphi 
        \label{eq:omega}
    \end{equation}
    where $\varphi$ is the azimuthal angle and $\theta$ is the direction window given by the user.
    \item Estimate the relative error for each brightness value by Eq.~\ref{eq:rerr}:
    \begin{equation}
        r.err = \frac{\sigma }{\sqrt{N}}
        \label{eq:rerr}
    \end{equation}
    in which $N$ is the total weight of the specified window and $\sigma$ is the standard deviation and calculated as Eq.~\ref{eq:sigma}:
    \begin{equation}
        \sigma =\sqrt{\frac{\sum_{i=1}^{N}(\frac{w_{i}}{\overline{w}})^{2}-N\overline{w^{2}}}{N-1}}
        \label{eq:sigma}
    \end{equation}
    where $\overline{w}$ is the mean weight of all the neutrons.

\end{itemize}

The output of Brightify will be a brightness map, showing the brightness for each position window and its corrresponding mean direction, represented by arrows. In this representation, the arrow length is proportional to $\sin\theta$, while its orientation reflects the azimuthal angle $\varphi$. Arrows with near-zero length indicate that the mean direction is nearly parallel to the beam direction (i.e., the z-axis).

\subsection*{Input Parameters}
In addition to the MCPL file, the user must specify six other input parameters:
\begin{enumerate}
    \item \textbf{Type of the recording surface (flat/curved)}: the user should determine the type of the surface which the MCPL file is created on. This tool offers the capability to calculate brightness on both flat and curved (spherical) surfaces. 
    \item \textbf{Type of the particle}: applicable to any particle recorded in the MCPL file (e.g., neutrons, protons).
    \item \textbf{Energy range}: selects particles in specified energy intervals.
    \item \textbf{Number of scan points} (only for the curved surfaces): defines coverage density and spatial resolution. A spiral or grid scan is used depending on the surface geometry \cite{bauer_distribution_2000, arthur_pdf_nodate}. This parameter is not required for the flat surfaces, as the number of scan points is automatically calculated based on the position window size.
    \item \textbf{Position window}: the area of the mesh in the phase~space called position window and it is in the units of cm$^2$.
    \item \textbf{Direction window}: the desirable divergence of the neutrons inside the phase~space mesh is called direction window and is in the units of degrees.
\end{enumerate}

\begin{figure}[!htb]
\centering
\includegraphics[width=0.7\columnwidth]{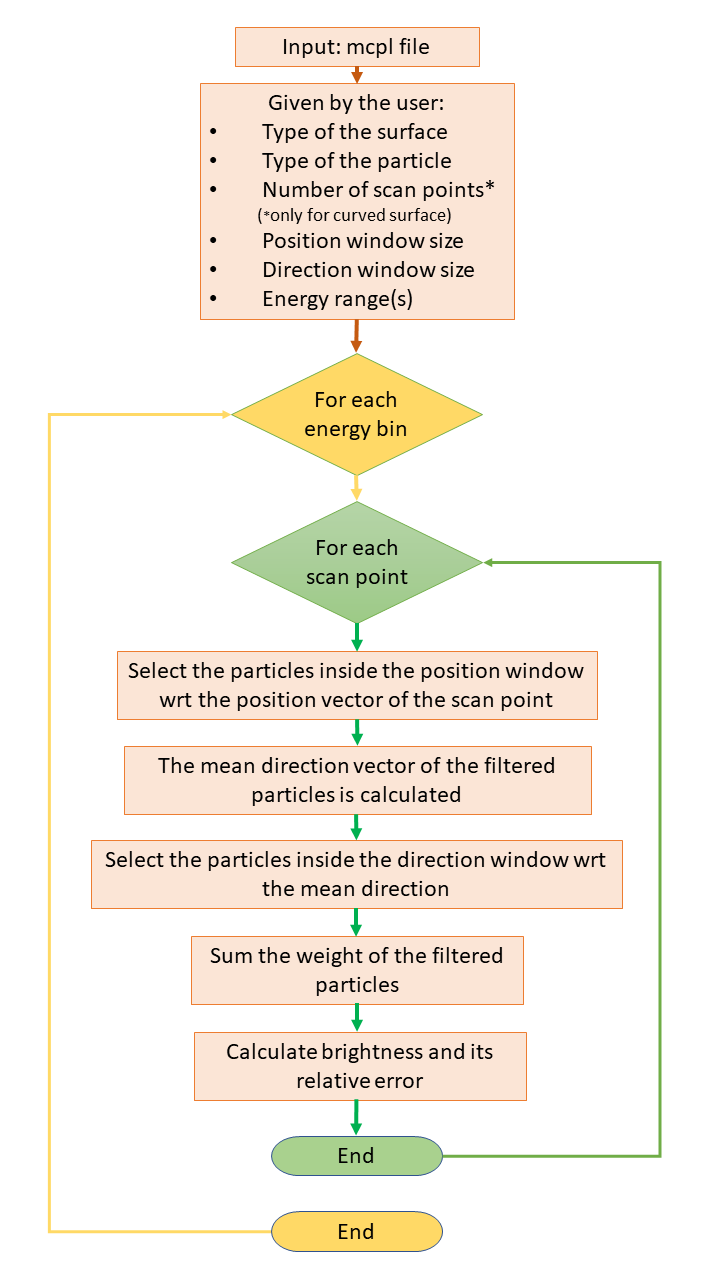}
\caption{The flowchart for the Brightify algorithm.}
\label{fig:brightify}
\end{figure}

Fig. \ref{fig:brightify} shows the flowchart for the Brightify algorithm.

Brightify is designed for efficient processing of large MCPL files using vectorized NumPy operations. While typical use cases (up to 10$^7$ particles) do not require parallelization, memory usage during MCPL file loading can become a limiting factor for very large datasets. In such cases, performance can be improved by applying data chunking strategies to reduce memory load. 

\begin{figure}[!htb]
\centering
\includegraphics[width=0.8\columnwidth]{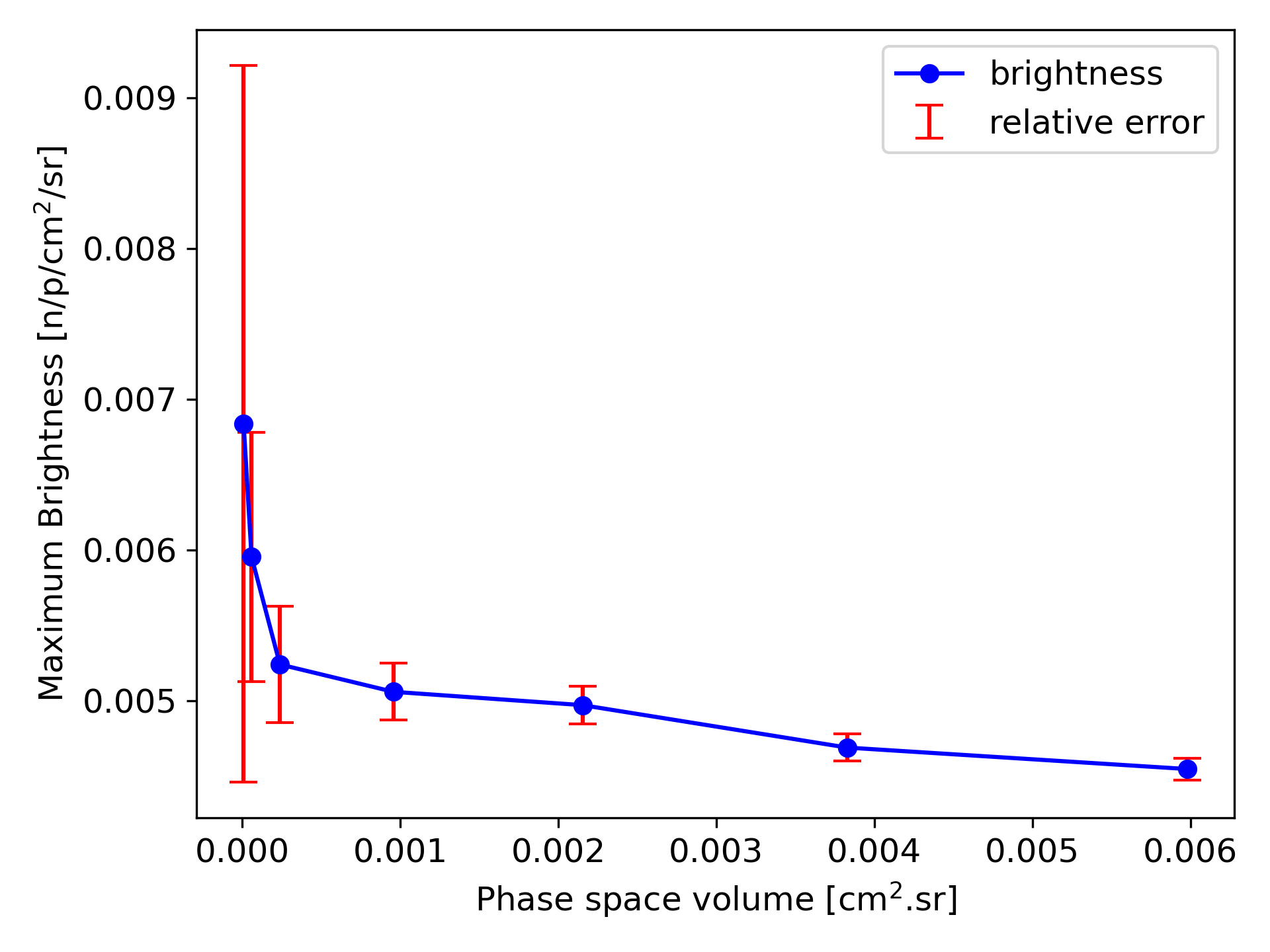}
\caption{Maximum brightness as a function of phase~space volume and the corresponding relative error, calculated by Brightify.}
\label{fig:brightify-r-err}
\end{figure}

The effectiveness of the brightness calculation in Brightify, like the other methods, is governed by two main factors: 
\begin{itemize}
    \item Neutron distribution function: This is determined by the MCPL input file, i.e. the discrete presentation of the neutron distribution function. This input can affect the calculation in two aspects: first, if an appropriate surface is chosen for the MCPL file. Second, if sufficient statistics is used for the MCPL file, as inadequate statistics can result in imprecise brightness values. 
    \item phase~space volume ($\Delta A \cdot \Delta \Omega$), which is determined by the size of position window and direction window in Brightify context. The size of these two windows will limit the number of neutrons inside the phase~space volume which directly affects the relative error. Too small phase~space volume leads to a high statistical error and consequently unreliable results, as illustrated in Fig.~\ref{fig:brightify-r-err}. On the other hand, if it is too large, the local brightness might average out. Therefore, special care must be taken when interpreting local peaks in brightness maps at high resolution. Balancing resolution and statistics, and validating local maxima against relative error thresholds (typically below 5\%~\cite{kulesza_mcnp_2022}), helps prevent over-interpretation of noise.
\end{itemize}

To have a full coverage and avoid gaps in the brightness map, it is advisable to overlap neighboring position windows by setting scan point spacing smaller than the position window size. This ensures that all neutrons contribute to the analysis and that local maxima are not missed due to undersampling.

\section{Verification and Validation}

To validate the performance of Brightify against the conventional methods—surface current tally and point estimator tally—we consider three test cases, each representing a different type of neutron distribution function:
\begin{itemize}
    \item (a) A uniform neutron distribution in both position $\vec{r}$ and direction $\hat{\Omega}$, using a flat isotropic source.
    \item (b) A symmetric distribution, modeled using a flat centric Gaussian source.
    \item (c) An asymmetric distribution, created by an off-center flat Gaussian source.
\end{itemize}

In all cases, the source emits 25 meV mono-energetic neutrons, and the simulations are performed in steady-state mode (i.e., integrated over time). Consequently, the calculated brightness in each case~is influenced by the phase~space volume, $\Delta A \cdot \Delta \Omega$.

For each configuration, we classify the phase~space volume based on whether the area $\Delta A$ is smaller than or comparable to the moderator surface area. The performance of Brightify is then compared with that of the surface current and point estimator methods accordingly.

\begin{figure}[!htb]
\centering
\includegraphics[width=1.0\columnwidth]{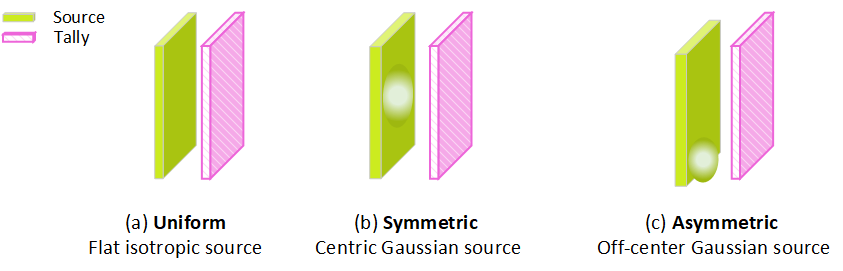}
\caption{Three test cases for validation of Brightify performance against conventional methods.}
\label{fig:validation-schematic}
\end{figure}

In each scenario, a flat surface is used for neutron recording. Fig.~\ref{fig:validation-schematic} presents schematics of the three configurations, with the recording surfaces for Brightify and surface current tallies indicated by dashed pink color and the source with green. For all cases, the point detector tally is positioned 23~cm from the center of the recording surface, inside an artificial collimator corresponding to the defined angular acceptance.

All Monte~Carlo simulations were performed using PHITS. In case~(a), a built-in surface source with dimensions of 2$\times$2~cm$^2$, oriented normal to the $z$ axis, was used. For cases (b) and (c), a built-in Gaussian source with a full width at half maximum (FWHM) of 0.5 cm in $x$ and $y$ dimensions was used. The tally surface in all three cases had the same dimensions of 2$\times$2 cm$^2$.

Table \ref{tab:validation-results} summarizes the results for a phase~space volume defined by a position window of $\Delta A$ = 4 cm$^2$ (equal to the tally surface area) and a direction window of $\Delta \Omega$~=~0.006~sr, which corresponds to a cone with a 5$^\circ$ opening angle.

\begin{table}[!htb]
\centering
\caption{Comparison of brightness values (in n/source/cm$^2$/sr) calculated using Brightify, surface current tally, and point estimator tally for the three test cases ($\Delta A = 4$\,cm$^2$ and $\Delta \Omega~=~0.006~$~sr).}
\label{tab:validation-results}
\resizebox{\columnwidth}{!}{%
\begin{tabular}{lccc}
\toprule
\textbf{Test Case} & \textbf{Brightify} & \textbf{Surface Current Tally} & \textbf{Point Estimator Tally} \\
\midrule
(a)        & $1.9838e-2 \pm 0.0067$  & $1.9840e-2 \pm 0.0067$ & $1.9922e-2$ \\
(b)     & $2.0090e-2 \pm 0.0066$ & $2.0123e-2 \pm 0.0066$ & $1.9987e-2$ \\
(c)     & $1.1475e-2 \pm 0.0088$ & $1.0098e-2 \pm 0.0094$ & $9.9593e-3$ \\
\bottomrule
\end{tabular}
}
\end{table}

As observed, for a large $\Delta A$—corresponding to a low-resolution brightness map—all three methods yield consistent results within the error margins for both the uniform source (case~a), where the neutron distribution is independent of position and direction, and the symmetric source (case~b), where the surface normal aligns with the mean direction in Brightify due to symmetry. 

However, the methods diverge in the asymmetric configuration (c), where the Gaussian source centroid is shifted downward and therefore tilting the peak direction by a few degrees relative to the tally surface normal. Both the surface‐current and point‐estimator tallies use a fixed 5$^\circ$ cone about that normal and thus systematically miss the tilted trajectories. Brightify, in contrast, computes the true mean direction of the neutron beam and recenters its 5$^\circ$ cone on that axis, recovering the otherwise‐lost flux. This 14 \% increase thus reflects only the geometric realignment of the acceptance cone, which vanishes whenever the beam and surface normal coincide.

\begin{figure}[!htb]
\centering
\includegraphics[width=1.0\textwidth]{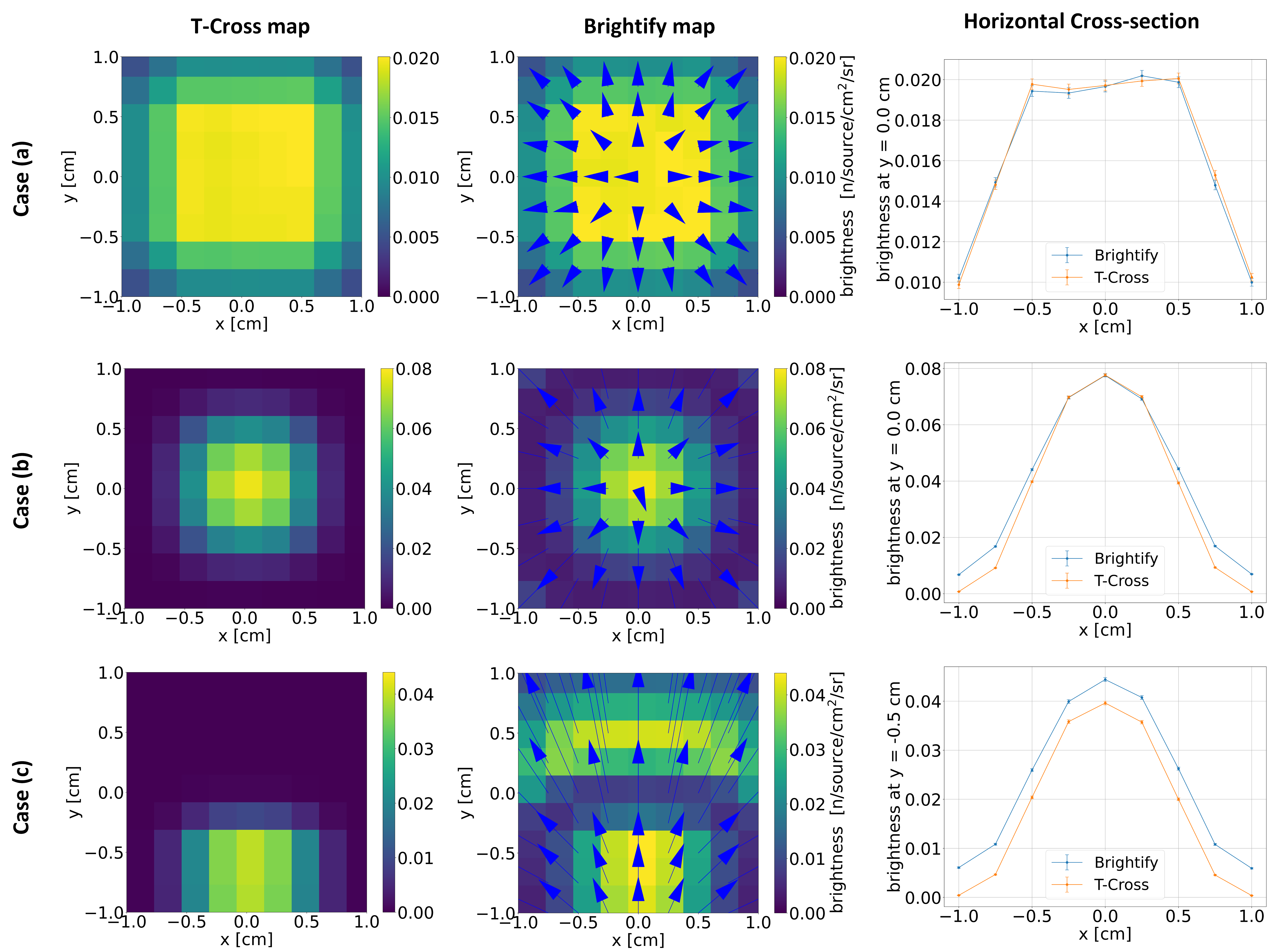}
\caption{Comparison of brightness maps obtained using the surface current tally and Brightify for the three test cases: (top) case~(a) — uniform source; (middle) case~(b) — symmetric source; (bottom) case~(c) — asymmetric source. In each panel, the brightness map from the surface current tally is shown on the left, the map from Brightify is in the center, and a horizontal cross-section along the axis of maximum brightness is presented on the right. For all cases, $\Delta A = 1$~cm$^2$ corresponds to the size of position window, and $\Delta \Omega = 0.006$~sr corresponds to the size of direction window.}
\label{fig:brightmap_comparison}
\end{figure}

In contrast, when instrument requirements demand a finer spatial resolution (i.e., smaller $\Delta A$ compare to the moderator area), the point estimator tally becomes impractical or overly complex to implement, due to definition of artificial collimators. The surface current tally can still be applied by defining an X–Y mesh on the recording surface, however with the limitation of brightness calculation only with respect to the normal direction.

Fig.~\ref{fig:brightmap_comparison} presents the brightness maps obtained using the surface current tally and Brightify, along with a cross-sectional comparison to highlight differences between the two approaches across all three test cases.

The blue arrows in brightness map by Brightify show the direction for which the brightness is calculated within each position window. For the uniform source (case~a) and the symmetric source (case~b), Brightify and the surface current tally show good agreement within the error margins when capturing the maximum brightness. However, in the asymmetric source (case~c), a clear increase in brightness of 10–15\% is observed when the brightness is calculated relative to the local mean directions. A similar gain is also noticeable in case~(b) at off-center positions.

The absolute brightness values reveal that, in case~(a), the brightness remains unchanged between the large phase~space volume (i.e., the larger area in Table~\ref{tab:validation-results}) and the smaller area shown in the top panel of Fig.~\ref{fig:brightmap_comparison}, which is expected for a uniform source. In contrast, cases (b) and (c) exhibit a fourfold increase in brightness when evaluated within the smaller phase~space area.

These gains are demonstrated using a simplified two-dimensional source with relatively uniform directional distribution and the gain factors can be different for realistic three-dimensional sources, depending on the relevant phase~space volume and the characteristics of the neutron source distribution. In scenarios where the neutron directions vary significantly within a position window, the mean direction may not fully represent the direction of peak brightness, and additional angular analysis could be necessary. It is important to note that in most practical cases, such as spallation sources or compact neutron sources, the neutron distribution is asymmetric. As such, these setups more closely resemble case~(c), where proper orientation is essential to achieve maximum brightness gain.

\section{Conclusions}

Monte~Carlo simulations are an essential technique for the design and optimization of neutron sources, yet they are often slow and computationally expensive. This makes iterative design processes particularly challenging. According to Liouville's theorem, brightness is conserved in conservative force fields along the particle trajectories, and serves as a key performance metric linking the source to the optics and a quantity to evaluate the performance of the source. However, calculating brightness directly within Monte~Carlo codes over different positions and directions remains complex and limited.

In this paper, we introduced \textit{Brightify}, an open-source Python tool designed to compute neutron brightness from MCPL files, which can be generated from various Monte~Carlo packages. Brightify offers a flexible, directionally-aware approach to brightness evaluation by scanning over spatial and angular domains, allowing users to evaluate brightness along mean directions at each position window, providing a practical representative metric while leaving room for future refinement in cases with complex directional distributions.

Through validation against standard methods, surface current tally and point estimator tally, we demonstrated that Brightify produces consistent results for large phase~space volumes while offering improved resolution and adaptability for finer analyses. In particular, Brightify overcomes the limitations of traditional methods, such as the directional constraints of surface current tallies and the impracticality of point tallies in close source-to-guide distance. Brightify is especially practical in compact accelerator-based neutron sources, where guides can be placed closer to the source.

By enabling localized and directionally-resolved brightness calculations, Brightify streamlines the evaluation and re-optimization of neutron sources. It offers researchers an accessible and extensible tool for post-processing Monte~Carlo outputs. 

Future versions of Brightify may include a dedicated module to improve robustness with low-statistics MCPL files, as well as support for parallel processing to better accommodate high-statistics simulations. An adaptive workflow may also be introduced to assist users in selecting an appropriate brightness map resolution. The full code is available on the Brightify GitHub repository \cite{akhyani_mina_brightify_nodate}.

\section{Acknowledgments}

This project has received funding from the European Union's Horizon 2020 research and innovation programme under the Marie Skłodowska-Curie grant agreement No. 754354 and No. 101034266.

%\appendix

% \section{Sample Appendix Section}
% \label{sec:sample:appendix}
% Lorem ipsum dolor sit amet, consectetur adipiscing elit, sed do eiusmod tempor section \ref{sec:sample1} incididunt ut labore et dolore magna aliqua. Ut enim ad minim veniam, quis nostrud exercitation ullamco laboris nisi ut aliquip ex ea commodo consequat. Duis aute irure dolor in reprehenderit in voluptate velit esse cillum dolore eu fugiat nulla pariatur. Excepteur sint occaecat cupidatat non proident, sunt in culpa qui officia deserunt mollit anim id est laborum.

\bibliographystyle{elsarticle-num} 
\bibliography{biblio}

\begin{thebibliography}{10}
\expandafter\ifx\csname url\endcsname\relax
  \def\url#1{\texttt{#1}}\fi
\expandafter\ifx\csname urlprefix\endcsname\relax\def\urlprefix{URL }\fi
\expandafter\ifx\csname href\endcsname\relax
  \def\href#1#2{#2} \def\path#1{#1}\fi

\bibitem{akhyani_mina_brightify_nodate}
M.~Akhyani, L.~Zanini, H.~M. Rønnow, \href{https://github.com/BrightnessTools/Brightify}{Brightify, a tool for calculating brightness in neutron sources}.
\newline\urlprefix\url{https://github.com/BrightnessTools/Brightify}

\bibitem{niita_phitsparticle_2006}
K.~Niita, T.~Sato, H.~Iwase, H.~Nose, H.~Nakashima, L.~Sihver, \href{https://www.sciencedirect.com/science/article/pii/S1350448706001351}{{PHITS}—a particle and heavy ion transport code system}, Radiation Measurements 41~(9) (2006) 1080--1090.
\newblock \href {https://doi.org/10.1016/j.radmeas.2006.07.013} {\path{doi:10.1016/j.radmeas.2006.07.013}}.
\newline\urlprefix\url{https://www.sciencedirect.com/science/article/pii/S1350448706001351}

\bibitem{forster_mcnp_1985}
R.~A. Forster, T.~N.~K. Godfrey, {MCNP} - a general {Monte} {Carlo} code for neutron and photon transport, in: R.~Alcouffe, R.~Dautray, A.~Forster, G.~Ledanois, B.~Mercier (Eds.), Monte-{Carlo} {Methods} and {Applications} in {Neutronics}, {Photonics} and {Statistical} {Physics}, Lecture {Notes} in {Physics}, Springer, Berlin, Heidelberg, 1985, pp. 33--55.
\newblock \href {https://doi.org/10.1007/BFb0049033} {\path{doi:10.1007/BFb0049033}}.

\bibitem{romano_openmc_2013}
P.~K. Romano, B.~Forget, \href{https://www.sciencedirect.com/science/article/pii/S0306454912003283}{The {OpenMC} {Monte} {Carlo} particle transport code}, Annals of Nuclear Energy 51 (2013) 274--281.
\newblock \href {https://doi.org/10.1016/j.anucene.2012.06.040} {\path{doi:10.1016/j.anucene.2012.06.040}}.
\newline\urlprefix\url{https://www.sciencedirect.com/science/article/pii/S0306454912003283}

\bibitem{agostinelli_geant4simulation_2003}
S.~Agostinelli, {others}, {GEANT4}–a simulation toolkit, Nucl. Instrum. Meth. A 506 (2003) 250--303.
\newblock \href {https://doi.org/10.1016/S0168-9002(03)01368-8} {\path{doi:10.1016/S0168-9002(03)01368-8}}.

\bibitem{battistoni_overview_2015}
G.~Battistoni, T.~Boehlen, F.~Cerutti, P.~W. Chin, L.~S. Esposito, A.~Fassò, A.~Ferrari, A.~Lechner, A.~Empl, A.~Mairani, A.~Mereghetti, P.~G. Ortega, J.~Ranft, S.~Roesler, P.~R. Sala, V.~Vlachoudis, G.~Smirnov, \href{https://www.sciencedirect.com/science/article/pii/S0306454914005878}{Overview of the {FLUKA} code}, Annals of Nuclear Energy 82 (2015) 10--18.
\newblock \href {https://doi.org/10.1016/j.anucene.2014.11.007} {\path{doi:10.1016/j.anucene.2014.11.007}}.
\newline\urlprefix\url{https://www.sciencedirect.com/science/article/pii/S0306454914005878}

\bibitem{lux_monte_2018}
I.~Lux, Monte {Carlo} {Particle} {Transport} {Methods}, CRC Press, 2018, google-Books-ID: y0paDwAAQBAJ.

\bibitem{gallmeier_source_2005}
F.~X. Gallmeier, Source {Terms} for {Neutron} {Beamline} {Shielding} and {Activation} {Calculations}, Tech. Rep. SNS-107030700-DA0002-R00, ORNL (Feb. 2005).

\bibitem{landau_lev_davidovich_course_1958}
L.~D. Landau, Course of {Theoretical} {Physics}, Vol.~5, Pergamon Press, 1958.

\bibitem{willendrup_mcstas_2014}
P.~Willendrup, E.~Farhi, E.~Knudsen, U.~Filges, K.~Lefmann, \href{https://content.iospress.com/articles/journal-of-neutron-research/jnr004}{{McStas}: {Past}, present and future}, Journal of Neutron Research 17~(1) (2014) 35--43, publisher: IOS Press.
\newblock \href {https://doi.org/10.3233/JNR-130004} {\path{doi:10.3233/JNR-130004}}.
\newline\urlprefix\url{https://content.iospress.com/articles/journal-of-neutron-research/jnr004}

\bibitem{lefmann_mcstas_1999}
K.~Lefmann, K.~Nielsen, \href{https://doi.org/10.1080/10448639908233684}{{McStas}, a general software package for neutron ray-tracing simulations}, Neutron News 10~(3) (1999) 20--23, publisher: Taylor \& Francis \_eprint: https://doi.org/10.1080/10448639908233684.
\newblock \href {https://doi.org/10.1080/10448639908233684} {\path{doi:10.1080/10448639908233684}}.
\newline\urlprefix\url{https://doi.org/10.1080/10448639908233684}

\bibitem{kulesza_mcnp_2022}
J.~A. Kulesza, T.~R. Adams, J.~C. Armstrong, S.~R. Bolding, F.~B. Brown, J.~S. Bull, T.~P. Burke, A.~R. Clark, R.~A.~A. Forster~III, J.~F. Giron, T.~S. Grieve, C.~J. Josey, R.~L. Martz, G.~W. McKinney, E.~J. Pearson, M.~E. Rising, C.~J.~C. Solomon~Jr., S.~Swaminarayan, T.~J. Trahan, S.~C. Wilson, A.~J. Zukaitis, \href{https://www.osti.gov/biblio/1889957}{{MCNP}® {Code} {Version} 6.3.0 {Theory} \& {User} {Manual}}, Tech. Rep. LA-UR-22-30006, Los Alamos National Lab. (LANL), Los Alamos, NM (United States) (Sep. 2022).
\newblock \href {https://doi.org/10.2172/1889957} {\path{doi:10.2172/1889957}}.
\newline\urlprefix\url{https://www.osti.gov/biblio/1889957}

\bibitem{zanini_design_2019}
L.~Zanini, K.~H. Andersen, K.~Batkov, E.~B. Klinkby, F.~Mezei, T.~Schönfeldt, A.~Takibayev, \href{https://www.sciencedirect.com/science/article/pii/S0168900219300087}{Design of the cold and thermal neutron moderators for the {European} {Spallation} {Source}}, Nuclear Instruments and Methods in Physics Research Section A: Accelerators, Spectrometers, Detectors and Associated Equipment 925 (2019) 33--52.
\newblock \href {https://doi.org/10.1016/j.nima.2019.01.003} {\path{doi:10.1016/j.nima.2019.01.003}}.
\newline\urlprefix\url{https://www.sciencedirect.com/science/article/pii/S0168900219300087}

\bibitem{ornl_2017}
T.~C. McClanahan, F.~X. Gallmeier, E.~B. Iverson, \href{https://www.osti.gov/biblio/1345791}{Moderator demonstration facility design and optimization}, Tech. rep., Oak Ridge National Lab. (ORNL), Oak Ridge, TN (United States). Spallation Neutron Source (SNS) (02 2017).
\newblock \href {https://doi.org/10.2172/1345791} {\path{doi:10.2172/1345791}}.
\newline\urlprefix\url{https://www.osti.gov/biblio/1345791}

\bibitem{bruckel_conceptual_nodate}
T.~Brückel, T.~Gutberlet, Conceptual {Design} {Report} {Jülich} {High} {Brilliance} {Neutron} {Source} ({HBS}).

\bibitem{batkov_unperturbed_2013}
K.~Batkov, A.~Takibayev, L.~Zanini, F.~Mezei, \href{https://www.sciencedirect.com/science/article/pii/S0168900213010073}{Unperturbed moderator brightness in pulsed neutron sources}, Nuclear Instruments and Methods in Physics Research Section A: Accelerators, Spectrometers, Detectors and Associated Equipment 729 (2013) 500--505.
\newblock \href {https://doi.org/10.1016/j.nima.2013.07.031} {\path{doi:10.1016/j.nima.2013.07.031}}.
\newline\urlprefix\url{https://www.sciencedirect.com/science/article/pii/S0168900213010073}

\bibitem{rizzi_intense_2024}
N.~Rizzi, B.~Folsom, M.~Akhyani, M.~Bertelsen, P.~Böni, Y.~Beßler, T.~Bryś, A.~Chambon, V.~Czamler, B.~Lauritzen, J.~I.~M. Damián, V.~Nesvizhevsky, B.~Rataj, S.~Samothrakitis, V.~Santoro, H.~Shuai, M.~Strobl, M.~Strothmann, A.~Takibayev, R.~Wagner, L.~Zanini, O.~Zimmer, \href{https://linkinghub.elsevier.com/retrieve/pii/S0168900224001414}{An intense source of very cold neutrons using solid deuterium and nanodiamonds for the {European} {Spallation} {Source}}, Nuclear Instruments and Methods in Physics Research Section A: Accelerators, Spectrometers, Detectors and Associated Equipment 1062 (2024) 169215.
\newblock \href {https://doi.org/10.1016/j.nima.2024.169215} {\path{doi:10.1016/j.nima.2024.169215}}.
\newline\urlprefix\url{https://linkinghub.elsevier.com/retrieve/pii/S0168900224001414}

\bibitem{kittelmann_monte_2017}
T.~Kittelmann, E.~Klinkby, E.~B. Knudsen, P.~Willendrup, X.~X. Cai, K.~Kanaki, \href{https://www.sciencedirect.com/science/article/pii/S0010465517301261}{Monte {Carlo} {Particle} {Lists}: {MCPL}}, Computer Physics Communications 218 (2017) 17--42.
\newblock \href {https://doi.org/10.1016/j.cpc.2017.04.012} {\path{doi:10.1016/j.cpc.2017.04.012}}.
\newline\urlprefix\url{https://www.sciencedirect.com/science/article/pii/S0010465517301261}

\bibitem{konik_new_2023}
P.~Konik, A.~Ioffe, \href{https://www.sciencedirect.com/science/article/pii/S0168900223006332}{A new method to find out the optimal neutron moderator size based on neutron scattering instrument parameters}, Nuclear Instruments and Methods in Physics Research Section A: Accelerators, Spectrometers, Detectors and Associated Equipment 1056 (2023) 168643.
\newblock \href {https://doi.org/10.1016/j.nima.2023.168643} {\path{doi:10.1016/j.nima.2023.168643}}.
\newline\urlprefix\url{https://www.sciencedirect.com/science/article/pii/S0168900223006332}

\bibitem{akhyani_mina_high_2024}
{Akhyani, Mina}, {Ronnow, Henrik M.}, {Zanini, Luca}, \href{https://doi.org/10.1051/epjconf/202429805001}{A high brilliance low-frequency spallation source optimized for one single instrument}, EPJ Web Conf. 298 (2024) 05001.
\newblock \href {https://doi.org/10.1051/epjconf/202429805001} {\path{doi:10.1051/epjconf/202429805001}}.
\newline\urlprefix\url{https://doi.org/10.1051/epjconf/202429805001}

\bibitem{bauer_distribution_2000}
R.~Bauer, \href{https://arc.aiaa.org/doi/10.2514/2.4497}{Distribution of {Points} on a {Sphere} with {Application} to {Star} {Catalogs}}, Journal of Guidance, Control, and Dynamics 23~(1) (2000) 130--137.
\newblock \href {https://doi.org/10.2514/2.4497} {\path{doi:10.2514/2.4497}}.
\newline\urlprefix\url{https://arc.aiaa.org/doi/10.2514/2.4497}

\bibitem{arthur_pdf_nodate}
M.~Arthur, \href{https://www.semanticscholar.org/paper/Point-Picking-and-Distributing-on-the-Disc-and-Arthur/10a2b95eaebf87f458db31f00aaf8c699c99caff}{[{PDF}] {Point} {Picking} and {Distributing} on the {Disc} and {Sphere} {\textbar} {Semantic} {Scholar}}.
\newline\urlprefix\url{https://www.semanticscholar.org/paper/Point-Picking-and-Distributing-on-the-Disc-and-Arthur/10a2b95eaebf87f458db31f00aaf8c699c99caff}

\end{thebibliography}





% \end{thebibliography}
\end{document}